\documentclass[twocolumn]{aastex631}

\newcommand{\angstrom}{\textup{\AA} }
\newcommand{\angstromns}{\textup{\AA}}
\newcommand{\nev}{\mbox{[Ne \textsc{v}] }}
\newcommand{\nevns}{\mbox{[Ne \textsc{v}]}}
\newcommand{\net}{\mbox{[Ne \textsc{iii}] }}
\newcommand{\ot}{\mbox{[O \textsc{iii}] }}

\newcommand{\of}{\mbox{[O \textsc{iv}] }}
\newcommand{\ofns}{\mbox{[O \textsc{iv}]}}
\newcommand{\netns}{\mbox{[Ne \textsc{iii}]}}

\newcommand{\netw}{\mbox{[Ne \textsc{ii}] }}
\newcommand{\netwns}{\mbox{[Ne \textsc{ii}]}}
\newcommand{\st}{\mbox{[S \textsc{iii}] }}
\newcommand{\stns}{\mbox{[S \textsc{iii}]}}
\newcommand{\sfo}{\mbox{[S \textsc{iv}] }}
\newcommand{\sfons}{\mbox{[S \textsc{iv}]}}

\newcommand{\netnetw}{\netns/\netw}
\newcommand{\ofnet}{\ofns/\net}

\newcommand{\nevnet}{\nevns/\net}

\usepackage{savesym}
\savesymbol{tablenum}
\usepackage{siunitx}
\restoresymbol{SIX}{tablenum}

\usepackage{amsmath}

\accepted{\today}

\submitjournal{ApJ}

\begin{document}

\title{Core Revelations: the Star Formation and AGN Connection at the Heart of NGC 7469}

\author[0000-0002-5718-2402]{Léa M. Feuillet}
\affiliation{Institute for Astrophysics and Computational Sciences and Department of Physics, The Catholic University of America, Washington, DC 20064, USA}

\author[0000-0003-4073-8977]{Steve Kraemer}
\affiliation{Institute for Astrophysics and Computational Sciences and Department of Physics, The Catholic University of America, Washington, DC 20064, USA}

\author[0000-0001-8485-0325]{Marcio B. Meléndez}
\affiliation{Space Telescope Science Institute, 3700 San Martin Drive Baltimore, MD
21218, USA}

\author[0000-0002-3365-8875]{Travis C. Fischer}
\affiliation{AURA for ESA, Space Telescope Science Institute, 3700 San Martin Drive, Baltimore, MD 21218, USA}

\author[0000-0003-2450-3246]{Henrique R. Schmitt}
\affiliation{Naval Research Laboratory, Remote Sensing Division, 4555 Overlook Ave SW, Washington, DC 20375, USA}

\author[0000-0002-0982-0561]{James N. Reeves}
\affiliation{Institute for Astrophysics and Computational Sciences and Department of Physics, The Catholic University of America, Washington, DC 20064, USA}

\author[0000-0001-8112-3464]{Anna Trindade Falcão}
\affiliation{Harvard-Smithsonian Center for Astrophysics, 60 Garden St., Cambridge, MA 02138, USA}



\begin{abstract}

We investigate the star formation-AGN connection in the Seyfert 1 NGC 7469 using James Webb Space Telescope (JWST) mid-infrared spectroscopic integral field unit (IFU) data. We use the IFU data to generate maps of different emission lines present in the spectrum, such as the star formation (SF) tracer \netw 12.81$\mu$m, and the AGN tracer \nev 14.32$\mu$m. We can separate the AGN- and SF-dominated regions using spatially resolved mid-infrared diagnostic diagrams, and further investigate the ionization sources powering each region by constructing photoionization models. We find that the previously detected eastern wind populates an intermediary region of the diagrams, between our star-forming and AGN points. Although it is possible that the star-forming ring may inherently not be uniform, this wind also coincides with a reduction in the \netw emission in the ring, which suggests that the ionization cone intersects the ring in this direction.

\end{abstract}



\section{Introduction} \label{sec:intro}

NGC 7469 is a nearby Seyfert 1.5 galaxy (69.4 Mpc away, or z = 0.01627), characterized by both narrow and broad components to H$\beta$ and other emission lines \citep{Veron2006}. Although it appears as a typical active galactic nucleus (AGN) in the optical, its infrared (IR) spectrum is dominated by the star formation in the host \citep{Weedman2005}. Notably, NGC 7469 also contains a powerful starburst ring located around the nucleus with a radius of approximately 500pc or 1.5'' \citep{Genzel1995}, making it an ideal target to explore the star-formation-AGN connection.  

The interplay between star formation and the presence of an AGN has been extensively studied using multi-wavelength data (e.g. \citealp{Vayner2017, Masoura2018, Feuillet2024-SFR}). However, with the advent of Integral Field Spectroscopy (IFS), we can now obtain highly detailed, spatially resolved grids of spectra (spaxels). The \emph{James Webb Space Telescope} (JWST) Mid-Infrared Instrument (MIRI), equipped with an integral field unit (IFU), offers an unprecedented opportunity to investigate the AGN-star formation connection within the inner 1 kpc region of NGC 7469 in the IR using the Medium Resolution Spectroscopy (MRS) mode.

NGC 7469 has been extensively analyzed in the past within various contexts. For instance, the MIRI study conducted by \cite{Lai2022} focused on the star-forming ring, analyzing dust through polycyclic aromatic hydrocarbon (PAH) emission and gas via pure rotational lines of H$_{2}$, calculating a star formation rate of 10-30 M$_{\odot}$yr$^{-1}$. Through their investigation, they found minimal impact of the AGN wind on the ring, similarly to the work of \cite{Zhang2023} which investigated the effect of the AGN on the warm molecular gas. The star-forming ring has been found to consist of two distinct stellar populations, a young (2-5 Myr) and an intermediate-age population ($\lesssim$30 Myr) \citep{Diaz-Santos2007, Bohn2023}.

The AGN itself has also been the focus of previous NGC 7469 studies. \cite{Armus2023} utilized MIRI’s high resolving power to examine the nucleus, identifying signs of an AGN-driven outflow. Other investigations of this galaxy's core region have detected the presence of an eastward wind originating from the central AGN. It has been detected by multiple instruments, including the \emph{Very Large Telescope}/Multi-unit Spectroscopic Explorer integral field spectrograph (MUSE IFS), and the MIRI and Near Infrared Spectrograph (NIRSpec) IFUs (\citealp{Robleto2021, Bianchin2024, U2022}). In particular, \cite{Bianchin2024} used NIR AGN shock diagnostics and found some evidence of shocked gas near the nucleus. However, the interaction between the AGN and the star formation in the ring warrants further investigation using mid-IR diagnostic diagrams.

Diagnostic diagrams use ratios of emission lines to differentiate between different ionizing sources, usually to separate AGN and star-forming galaxies among large spectroscopic data sets. The most common of them are known as Baldwin, Philip, \& Terlevich (BPT; \citealp{BPT1981}) and Veilleux \& Osterbrock (VO; \citealp{Veilleux1987}) diagrams and uses the \mbox{\ot 5007\angstromns/H$\beta$} vs either \mbox{[N \textsc{ii}] 6583\angstromns/H$\alpha$}, \mbox{[S \textsc{ii}] 6716\angstromns+5731\angstromns/H$\alpha$} or \mbox{[O \textsc{i}] 6300\angstromns/H$\alpha$} to distinguish between different sources of ionization in the optical. However, star-forming indicators in the optical are limited in number, affected by extinction, and often require deconvolution (e.g. [N \textsc{ii}] and H$\alpha$). Alternative options for these diagnostic diagrams have been developed in order to mitigate these issues (e.g. \citealp{Stasinska2006, Lamareille2010, Juneau2011, TBT2011, Yan2011, Juneau2014, Backhaus2022, Feuillet2024-Diagrams}). However, highly reliable star-forming indicators are available in the IR, such as \netw 12.81$\mu$m (hereafter \netwns, \citealp{Melendez2008-contribution}) and the aforementioned PAHs \citep{Sani2010}.

Both optical and IR diagnostic diagrams most often use integrated emission line values, as spectroscopic information commonly covers the full extent of the galaxy, such as data obtained from the \emph{Spitzer Space Telescope} \citep{Weaver2010}. Fully spatially resolved infrared diagnostic diagrams have been difficult to obtain as a result of the absence of any existing mid-IR IFUs. However, as previously stated, using MIRI IFS data allows us to obtain spatially resolved spectra and thus separate the AGN and star-formation-dominated regions within the core of NGC7469.

This paper is outlined as follows. Section \ref{sec:data} outlines the details of the observations and the data processing. Section \ref{sec: diags} demonstrates the effectiveness of using mid-IR diagnostic diagrams to separate the star-forming and AGN regions in NGC 7469. Section \ref{sec: models} outlines the different models that were run to further investigate the physical basis of the diagrams. Finally, we discuss our findings, specifically the origin and extent of AGN feedback on the star-forming ring based on our analysis in Section \ref{sec: discussion}. We use \emph{Wilkinson Microwave Anisotropy Probe} (WMAP) 12-year results cosmology throughout, with $H_0$ = 70.0 $km s^{-1} Mpc^{-1}$.

\section{Observations and Data Reduction}\label{sec:data}

The uncalibrated data was retrieved from the public Barbara A. Mikulski Archive for Space Telescopes (MAST). The observations of NGC 7469 were taken using the MIRI Medium Resolution Spectrometer as part of a Director's Discretionary Time (1328; PIs: L. Armus, A. S. Evans) Early Release Science (ERS) program \citep{Lai2022, U2022, Armus2023}. The data consists of spatially resolved spectra ranging from 4.9 to 27.9 $\mu m$ over a field of view (FOV) up to \SI{6.6}{\arcsecond}$\times$\SI{7.7}{\arcsecond} across four IFUs. Each IFU (designated as channels 1 through 4) has three grating settings: short, medium, and long. The default FASTR1 readout pattern was used, with an exposure time of 444 seconds. Background observations were also conducted, as the galaxy is considered an extended source.

The four available channels do not cover the same FOV, nor do they have the same number or size of spaxels by default. Given that a spaxel-to-spaxel comparison is essential to the aim of this paper, the data cubes must be convolved to match the plate scale of lowest resolution channel (channel 4). This process involves using the raw uncalibrated data, the background observations, and the MIRI pipeline (v1.16) to produce calibrated data cubes with uniform spaxel size and the same RA and DEC centering\footnote{The arguments in the Spec3 pipeline that were specified are the following: \emph{cube$\_$build.scalexy} = 0.273'', \emph{cube$\_$build.ra$\_$center} = 345.8151, \emph{cube$\_$build.dec$\_$center} = 8.8739.}.

\begin{figure*}[]
\centering
\includegraphics[width=\textwidth]{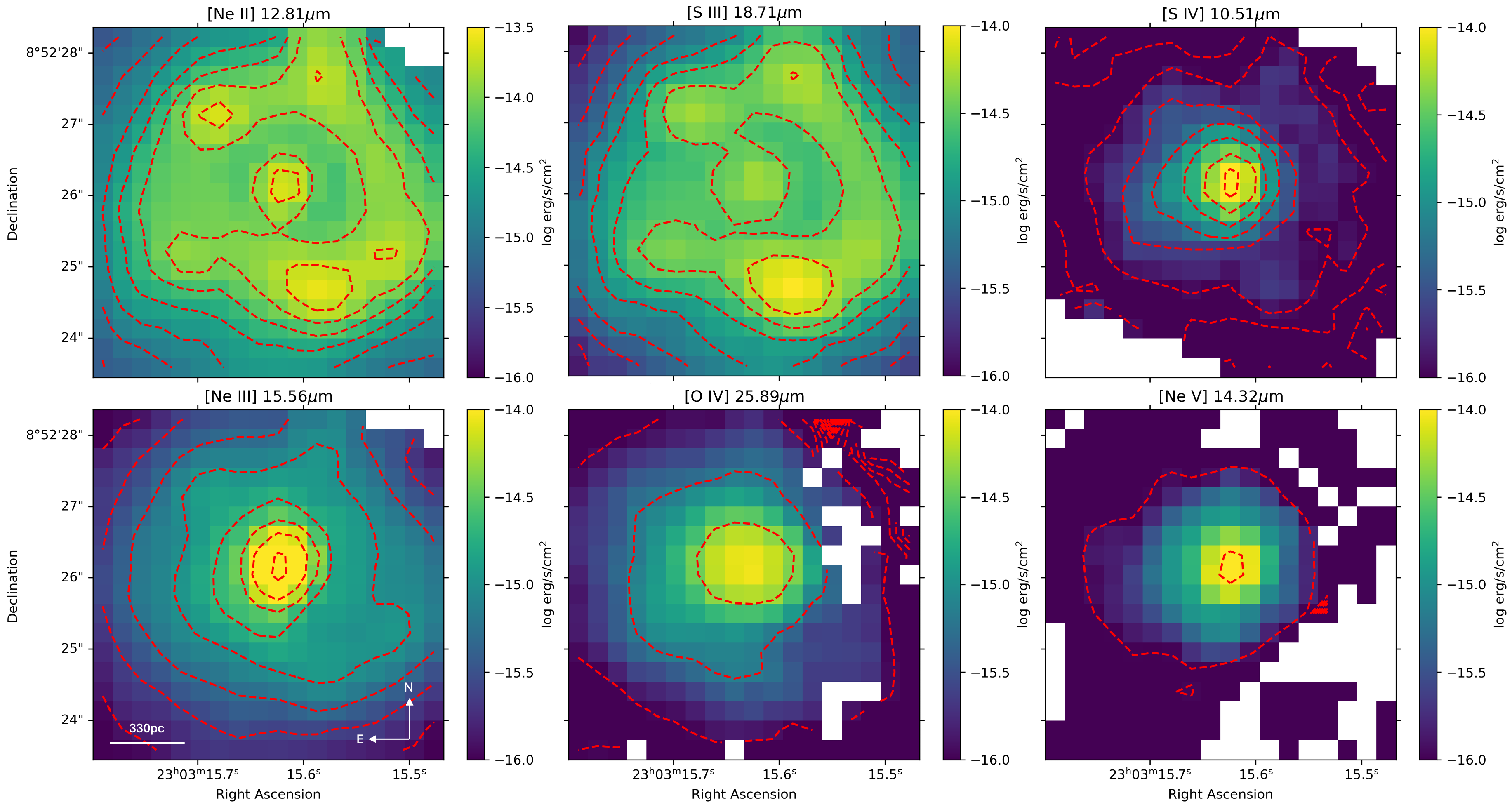}
    \caption{Maps for the emission lines used in the mid-IR diagnostic diagrams, in order of increasing IP: \netwns, \stns, \sfons, \netns, \ofns, and \nevns. The emission lines in the blank spaxels on the maps were below our signal-to-noise threshold and were therefore excluded. Contours have been added to the plots to emphasize the flux distribution structures. The star-forming ring is visible in the lower ionization line maps of \netw and \stns, as well as a decrease in their fluxes towards the east, creating a hole in the ring. The other maps also show some asymmetry in the flux distribution around the nucleus, where they extend further towards the east.  
    \label{fig: maps}}
\end{figure*}

The maps shown in Figure \ref{fig: maps} were obtained by fitting the emission line and the local continuum in each spaxel with up to three Gaussians and a linear model respectively by using the spectroscopic analysis tool PySpecKit \citep{Pyspeckit2011, Pyspeckit2022}. We impose a S/N threshold of 3 for all emission lines, with the exception of [Ne V] and [O IV], for which we use a cutoff of 2.5. The flux values in each spaxel were then derived by integrating the total modeled line. The maps cover a \SI{5}{\arcsecond}$\times$\SI{5}{\arcsecond} region centered on the nucleus. With consistent spaxel size and alignment, they allow for the creation of ratio maps, which are subsequently used to generate our spatially resolved mid-IR diagnostic diagrams presented below.



\section{Analysis}

\subsection{Mid-IR Diagnostic Diagrams} \label{sec: diags}

The wide wavelength coverage of the MIRI MRS instrument allows access to a multitude of emission lines that we may use as diagnostics in the mid-IR. The \net 15.56$\mu$m/\netw ratio is an effective option to separate AGN- and stellar-dominated regions. While \netw is a known star formation rate indicator \citep{Cao2008}, \net 15.56$\mu$m (hereafter \netns) is characteristic of AGN ionized gas, similar to \ot 5007\angstrom in the optical due to the ions that are produced from having similar ionization potentials (IPs, 41.0eV and 35.1eV respectively). However, \netw can also be produced in AGN-ionized gas, though in such cases, the \netnetw ratio tends to be greater than unity \citep{Melendez2008-contribution}. Consequently, a smaller ratio suggests a stronger contribution from star formation. We therefore expect the star-forming ring in NGC 7469 to have lower \netnetw values compared to the nucleus. 

All mid-IR diagnostic diagrams presented in this paper use the \netnetw ratio on the y-axis, with different ratios on the x-axis. An effective option uses \netnetw plotted against \of 25.89$\mu$m/\net \citep{Weaver2010}, but its use is limited in this case as \of faces both PSF and fringing issues in our data (see Appendix \ref{app: PSF} for more information). However, other MIR diagnostic diagrams are available, such as those using the 
\begin{itemize}
    \item \mbox{\sfo 10.51$\mu$m/\st 18.71$\mu$m},
    \item \mbox{\sfo 10.51$\mu$m/\net 15.56$\mu$m},
    \item \mbox{\nev 14.32$\mu$m/\net 15.56$\mu$m}
\end{itemize}
ratios. Like \ofns, a well-established AGN tracer \citep{Melendez2008-indicators}, \sfo and \nev also serve as AGN tracers because they originate from ions with relatively high ionization potential (IP) values of 34.8 eV and 97.1 eV, respectively\footnote{Interestingly, those ions that are characteristic of star formation have IP values below that of He$^0$, whereas those associated with AGN have IP values greater than that of He$^+$. Those with IPs between the two can be produced in either.}.

\begin{table}[]
\centering
\caption{Summary of the lines and fluxes}
\label{tab:lines}
\begin{tabular}{lccccc}
\hline
Line              & $\lambda_{rest}$ & IP  & n$_c$   & Flux$^{d}_{nucleus}$ & Flux$_{total}$ \\

& ($\mu$m) & (eV) & (cm$^{-3}$)& \multicolumn{2}{c}{10$^{-15}$ erg cm$^{-2}$ s$^{-1}$} \\
\hline\hline
\multicolumn{5}{l}{Channel 2} \\
\hline      
{[}S \textsc{iv}{]}        & 10.51     & 34.8 & 5.4e4$^a$ & 40.4 $\pm$ 2.0            & 107 $\pm$ 2.8          \\
\hline
\multicolumn{5}{l}{Channel 3}    \\
\hline
{[}Ne \textsc{ii}{]}       & 12.81     & 21.6 & 7.1e5$^b$ & 78.8 $\pm$ 4.1         & 2451 $\pm$ 45           \\
{[}Ne \textsc{v}{]}        & 14.32     & 97.1 & 3.5e4$^b$  & 57.9  $\pm$ 4.5           & 143 $\pm$ 1.1           \\
{[}Ne \textsc{iii}{]}      & 15.56     & 41.0 & 2.1e5$^b$  & 70.1 $\pm$ 2.8           & 426 $\pm$ 24           \\
\hline
\multicolumn{5}{l}{Channel 4}     \\
\hline
{[}S \textsc{iii}{]}       & 18.71     & 23.3 & 2.2e4$^a$ & 19.1 $\pm$ 1.2  & 890 $\pm$ 12           \\
{[}O \textsc{iv}{]}        & 25.89      & 54.9 & 5.0e3$^c$ & 43.6 $\pm$ 2.5   & 256 $\pm$ 12  \\
\hline
\end{tabular}
\tablecomments{IP is the ionization potential of the ion. n$_c$ is the critical density of the line, corresponding to the specified transition. $^a$Values obtained from Table 1 in \cite{Giveon2002}. $^b$Values obtained from Table 3.15 in \cite{Osterbrock2006}. $^c$Value obtained from \cite{Melendez2008-indicators}. $^d$Values are calculated from our defined inner nuclear region.}
\end{table}

We identified and fit the emission lines relevant to our work in the 1d galaxy spectrum output by the MIRI pipeline to obtain total flux values. The emission line fluxes calculated for the full field of view, as well as for what we define as our inner nucleus region\footnote{The nuclear values for the Channel 4 lines are likely underestimated due to PSF issues (see Appendix \ref{app: PSF}, \citealp{Law2023}).}, corresponding to the 4 brightest central spaxels in the [Ne II] map (\SI{0.5}{\arcsecond}$\times$\SI{0.5}{\arcsecond}), are given in Table \ref{tab:lines}. As can be seen in both the flux maps and the flux values in Table \ref{tab:lines}, both the \netw and \st appear prominently in the star-forming ring, while the AGN indicators \ofns, \nevns, and \sfo are all concentrated around the nucleus.

However, as the \st and \netw fluxes distinctly show the morphology of the clumpy star-forming ring, we can see that there appears to be a lack of emission coming from the east side of the ring (left side on the maps in Figure \ref{fig: maps}). This reduced emission is in the direction of the eastward outflowing wind from the AGN, as previously detailed. In particular, the highest ionization \of and \nev maps and contours show the eastward extension of the gas, which once again suggests the presence of this wind.

\begin{figure*}[]
\centering
\includegraphics[width=\textwidth]{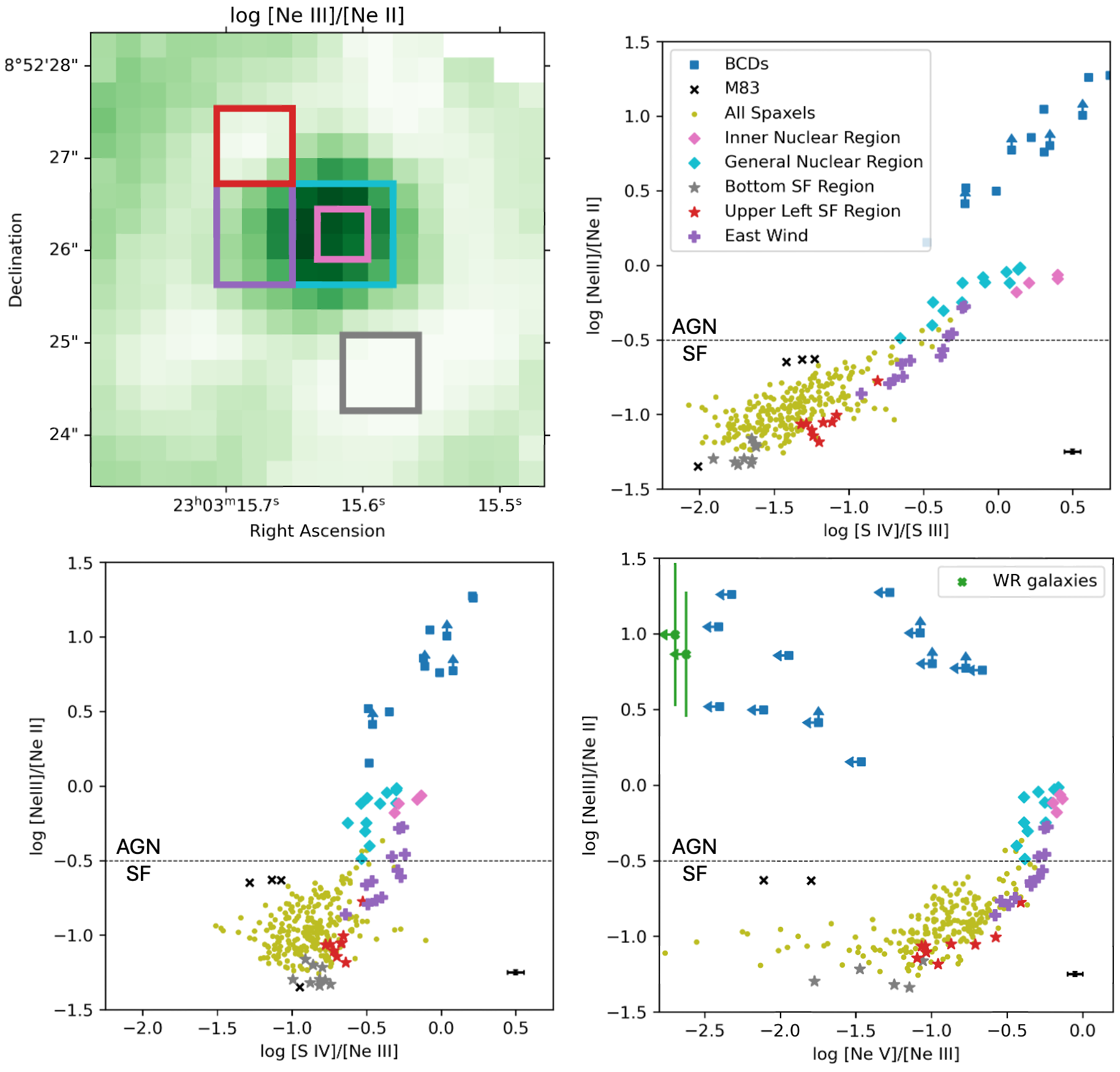}
    \caption{Spatially resolved diagnostic diagrams. Top left panel: map of the \netnetw ratio, with the nucleus and the star-forming ring showing different values. The other three panels show the mid-IR diagnostic diagrams, using \sfons/\stns, \sfons/\netns, and \nevns/\netns. The small khaki points each correspond to a spaxel on the map. The diamonds, stars, and pluses correspond to AGN-dominated, SF-dominated, the east wind spaxels respectively, with the different colors corresponding to spaxels of interest contained within the outlined regions on the map. The colors are as follows: the inner nucleus (pink), the general nuclear region (cyan), the east wind (purple), and the two most prominent clumps in the ring as seen on the \netw map (red in the north-east and grey in the south). Uncertainties in our ratio values were calculated on a spaxel-by-spaxel basis and then averaged to give a typical error bar (bottom right of diagrams). We also plot the AGN/SF demarcation line, which corresponds to the lowest [Ne III]/[Ne II] value that can be obtained from our AGN models in Section \ref{sec: models}.
    The blue squares are blue compact dwarf galaxies \citep{Hao2009-BCD}, the green crosses are Wolf-Rayet (WR) galaxies \citep{Goulding2009}, and the black crosses are MIRI/MRS data for M83, a nearby purely star-forming galaxy \citep{Hernandez2023}. 
    \label{fig: diagnostic diagrams}}
\end{figure*}

As mentioned in the previous section, we calculate the value for the flux of each emission line in each spaxel. Using these flux maps, we can then obtain ratio maps, such as \netnetw which is common across all three diagnostics. In the first panel of Figure \ref{fig: diagnostic diagrams}, we can see the \netnetw map, with the starburst ring displaying lower values as expected. We also highlight regions of interest, specifically the inner nucleus (pink), the general nuclear region (surrounding spaxels of the inner nucleus, where the region is distinctly AGN-dominated, cyan), the east wind (purple), and the two most prominent clumps in the ring as seen on the \netw map (red in the north-east and grey in the south).

The other three panels of Figure \ref{fig: diagnostic diagrams} show the mid-IR diagnostic diagrams that were previously discussed, with each khaki point corresponding to a single spaxel on the map. The star-, diamond-, and plus-shaped points are spaxels within the regions of interest with the same colors as defined in the previous paragraph. It is immediately evident that the nuclear points separate from the star-forming points quite clearly in all three diagrams. We also add a demarcation line between the AGN and SF regions of the diagram, which is defined by the lowest \netnetw ratio given by the AGN models in Section \ref{sec: models}. The east wind, which is in the direction of the hole in the ring, appears in an intermediate position between the AGN- and SF-dominated points. The other points present in the diagrams are outlined below: 

\begin{enumerate}
    \item The Blue Compact Dwarf (BCD) values are obtained from \cite{Hao2009-BCD}, which is used as a comparison within the context of mid-IR diagnostics (shown as blue squares). BCDs are small compact galaxies containing a large population of massive young stars, leading to their blue color. They appear at high \netns/\netw values as the gas is overionized, leading to an excess of high-ionization gas emitting lines such as \netns, but is unable to produce much \netwns. The arrows indicate upper or lower limits to the values. The BCDs do not overlap with any of the star-forming ring or AGN points. This indicates a significant difference in their ionization source compared to what is found in NGC 7469. Although both M83 and NGC 7469 have been shown to contain a significant number of Wolf-Rayet (WR) stars \citep{Hadfield2005-WR_M83, Miralles2016-WR_NGC7469}, we note the absence of any of our data points close to the BCDs or to the WR galaxies in the \nevnet diagram \citep{Goulding2009}. Assuming that these stars are indeed present, the contribution of neighboring O stars must be so significant as to drive everything firmly within the star-forming region (by increasing the \netw contribution).
    \item We additionally compare the results of the mid-IR diagnostic diagram for NGC 7469 to the purely star-forming galaxy M83, which was also observed by JWST using the MIRI instrument \citep{Hernandez2023}. This allows us to compare the placement of the intense star formation happening in the center of M83 to the star-forming ring in NGC 7469. Each point in the diagnostic diagrams corresponds to the average ratio found in four different regions of the core of the galaxy. The values obtained for M83 appear largely consistent with the location of our points located on the star-forming ring (see Figure \ref{fig: diagnostic diagrams}). The exception is found with the \nevnet diagnostic, with two of the four regions lacking any \nev detection, and the regions with detectable values having lower \nevnet values. This follows from the fact that \nev can only be made in the presence of an AGN, due to Ne$^{+4}$'s high IP. Our star-forming ring ratio values being so high compared to the M83 points indicates possible contamination from the AGN's ionizing radiation. 
\end{enumerate}

\subsection{Cloudy Modeling} \label{sec: models}

\begin{figure*}[]
\centering
\includegraphics[width=\textwidth]{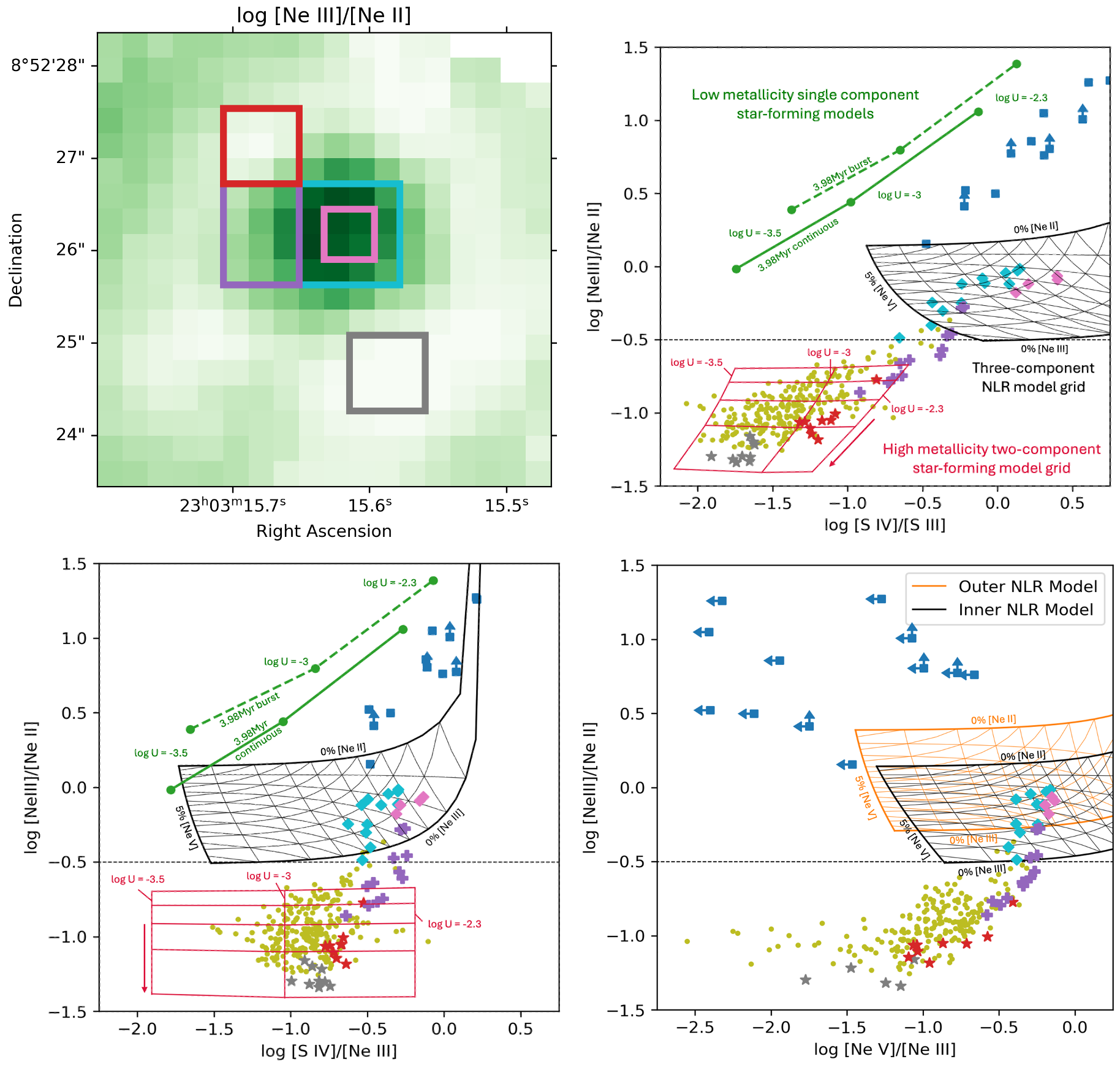}
    \caption{Spatially resolved diagnostic diagrams with photoionization model grids. The green tracks correspond to low metallicity (Z = 0.004) Starburst 99 models, with the solid track corresponding to the 3.98 Myr continuous star formation and the dashed track representing an instantaneous burst of star formation of the same age. A two-component star-forming model grid, with an older (10 Myr) and a younger (3.98 Myr) burst component is shown in red, with varying contribution of the older burst from 90 to 98\%, increasing along the arrow. Three-component NLR models are overlaid on the diagram in black, comprised of a very low ionization \netw component, a low ionization \net and \st component, and a high ionization \nev component. The tracks with the lowest contribution of each component are indicated on the plot, and the inner tracks correspond to increasing contributions from each track in 10\% increments. The NLR grids in black are generated from the inner models, while the orange one present on the bottom right diagram used the outer models. The lack of star-forming models on the \nevnet diagrams reflects the inability of these models to produce any significant amount of \nevns. 
    \label{fig: models}}
\end{figure*}

To compare our above data to theoretical models, we use the photo-ionization modeling code Cloudy (v. 23.01, \citealp{Cloudy2023}) to further investigate the physical conditions able to produce the various ratios used in the MIR diagnostic diagrams. We first adopt a narrow line region (NLR) model to reflect the conditions in the nucleus and construct a spectral energy distribution (SED) tailored to our galaxy based on previous UV and X-ray data of NGC 7469. We specifically use the values given in \cite{Scott2005} obtained using the \emph{Far Ultraviolet Spectroscopic Explorer} (FUSE) and the \emph{X-ray Multi-Mirror Mission} (XMM-Newton). We additionally use the method described in \cite{Kraemer1994}, which allows us to use the He II $\lambda$4686/H$\beta$ ratio to find the index of the missing section of the data in the extreme ultra-violet (13.6eV $\leq$ h$\nu$ $\leq$ 500eV), with the value of the ratio for NGC 7469 obtained from \cite{Cohen1983}. Our resulting broken power-law SED can be described as follows: 
\begin{equation*}
    \alpha = -1.0 \textrm{ for } 1 \textrm{ eV}  \leq \textrm{h}\nu < 4.96 \textrm{ eV;}
\end{equation*}
\begin{equation*}
        \alpha = -0.92 \textrm{ for } 4.96 \textrm{ eV}  \leq \textrm{h}\nu \leq 13.6 \textrm{ eV;}
\end{equation*}
\begin{equation*}
        \alpha = -1.67 \textrm{ for } 13.6 \textrm{ eV}  \leq \textrm{h}\nu \leq 500 \textrm{ eV;}
\end{equation*}
\begin{equation*}
        \alpha = -0.80 \textrm{ for } 500 \textrm{ eV} \leq \textrm{h}\nu \leq 100 \textrm{ keV.}
\end{equation*}

We further explore the validity of this SED by calculating the bolometric luminosity value obtained from the integration of the scaled SED, using the flux at 1000\angstrom given in \cite{Scott2005} as reference. The resulting bolometric luminosity equals $\log~L_\mathrm{bol (SED)}$ = 44.2 erg/s, which we then compare to bolometric luminosities obtained from the \ot 5007\angstrom luminosity \citep{Peterson2014} and the \cite{Lamastra2009} correction value of 454, $\log~L_\mathrm{bol([O  \textsc{iii}])}$ = 44.2 erg/s; from the 2-10 keV luminosity, $\log~L_\mathrm{bol(2-10keV)}$= 44.5 erg/s \citep{Vasudevan2007-2-10-bol-lum}; and from our value for the flux of \nev 14.32$\mu$m from \cite{Satyapal2007} (Equation 1), $\log~L_\mathrm{bol([Ne  \textsc{v}])}$ = 44.7 erg/s. We find good agreement between our SED-based bolometric luminosity\footnote{Using the SED-based bolometric luminosity and the SMBH mass value from \cite{Lu2021}, we calculate an Eddington ratio of 0.125$^{+0.025}_{-0.011}$ for NCG 7469's central black hole.} and the data-based ones, indicating that the models with this input SED will produce well-constrained values.  

We define the ionization parameter as
\begin{equation}\label{eq: U}
    U=\frac{Q}{4 \pi c r^{2} n_{H}}
\end{equation}
where Q is the number of ionizing photons per second emitted by the central object (based on our SED, \mbox{Q = $1.24 * 10^{54}$ photons s$^{-1}$)}, r is the distance between the central object and the cloud, c is the speed of light, and n$_{H}$ is the hydrogen number density \cite{Osterbrock2006}. We additionally use solar abundances, with exact values of C = -3.54, N = -4.17, O = -3.31, Ne = -3.94, Na = -5.78, Mg = -4.45, Al = -5.57, Si = -4.49, S = -4.88, Ar = -5.7, Ca = -5.7, Fe = -4.54, and Ni = -5.8, relative to hydrogen on a logarithmic scale \citep{Asplund2021}. We did not include dust in the models as none of the elements considered in this paper would be depleted onto dust, save for oxygen, and the effect would thus be minimal for the ratios we are investigating. We set the stopping criterion for the NLR models as reaching a column density value of 10$^{21.5}$, which is a typical value for the NLR \citep{Kraemer2000}. 

As we are dealing with spatially resolved data, we may tailor our models to the situation in a more precise way than has been done in the past. We first assumed a \netw component with log U = -4.5 and calculated the density range between the nucleus and the inner edge of the ring \citep{Melendez2014}. We used the approximate size of a single pixel for the inner component distance of 100pc, and a distance of 300pc for the outer component. Then, using Equation \ref{eq: U} and a Q value obtained from the previously stated SED we calculate the corresponding density range. Our NLR models use three different components: the \netw component outlined above with log U = -4.5 and two characteristic densities log n$_H$ = 6.04 and 5.08, a \net and \st component with log U = -3.5, and a \nev component with log U = -1. We assume that the gas is co-located, and thus vary the density for each component proportionally to the difference in ionization parameter, as suggested by Equation \ref{eq: U}. The resulting multi-component model grid is then obtained by varying the contribution of each component between 0\% and 100\% and shown in black in Figure \ref{fig: models}. One exception is made for the \nev component, which starts with a 5\% contribution at the low end, as the 0\% track appeared completely off the \nevnet diagram due to it being the only component able to make any significant amounts of \nevns. In each diagram, the inner NLR grid encompasses all of the nuclear points and the points directly adjacent to the nucleus, as well as a portion of the east wind points, while avoiding any star formation-dominated spaxels. The outer NLR grid is shown on the bottom right panel of Figure \ref{fig: models}, and has higher \netnetw values compared to the inner NLR grid, but a similar range of \nevns/\netns. This same behavior is seen on the other diagnostic diagrams (not shown here to reduce crowding of the diagrams). 
  
We then use Starburst99 models\footnote{\url{https://www.stsci.edu/science/starburst99/docs/parameters.html}} and Cloudy to predict the necessary ratios in star-forming galaxies \citep{Leitherer1999}. Our Starburst99 simulations use either a continuous star formation rate (SFR) of 20.0 \(M_\odot\)yr$^{-1}$ or an instantaneous burst with a total mass of 1$\times 10^{6} M_{\odot}$, and a Kroupa initial mass function (IMF) power law in both cases. We use the Padova with thermally pulsating asymptotic giant branch (AGB) track, as well as two different metallicities: Z = 0.004, 0.020. Any other parameter is left to the default values pre-filled in Starburst99. The Cloudy model assumes a variety of ages discussed below, a closed and spherical geometry, constant gas pressure, the temperature is solved for, and the density is set to log n$_{H}$ = 2.5. We vary the ionization parameter between log U = -3.5 and -2.3. The upper limit to log U for the HII region stems from the wind pressure dominating gravity as U increases, which creates winds that push the gas out from the HII region \citep{Yeh2012}.

The low metallicity single-component model tracks are shown in green in the top right and bottom left panels of Figure \ref{fig: models}, and do not overlap with our star-forming points in the ring. However, these tracks appear closer to the BCD and WR galaxies compared to the regular HII regions and the points in our ring, due to their characteristic lower metallicity values \citep{Kunth2000}. The tracks use an age of 3.98Myr, which allows enough time for WR stars to contribute to the models, as these are necessary to obtain high \netnetw ratios, such as the ones seen in BCD galaxies.

To obtain a model grid fitting to our ring data, we created a two-component model, with an older (10 Myr) and a younger burst component (3.98 Myr). We vary the contribution of the older burst from 90 to 98\% of the total H$\beta$ emission output by the models. The model grid is once again shown in the top right and bottom left panels of Figure \ref{fig: models}, with increasing contribution of the 10 Myr component following the arrow, and overlaps significantly with our star-forming points. This range of models is consistent with the findings of \cite{Diaz-Santos2007} and \cite{Bohn2023}, which shows two distinct populations in the ring of 2-5 Myr and 10-30 Myr respectively. 
The star-forming models do not produce any significant \nev and are therefore not included in the \nevnet diagram. The presence of \nev in the star formation-dominated regions requires an additional input of high energy radiation. This could be due to the influence of the AGN or due to the radiation resulting from radiative shocks with velocities $\sim$100s km/s, the latter due to supernovae \citep{Izotov2012}. However, as our NLR model grid successfully produces the \nevnet ratios in the AGN-dominated areas, there is no evidence for shock ionization in those regions.


\section{Discussion} \label{sec: discussion}

The primary objective of this paper is to explore the interaction between star formation and AGN activity at the core of NGC 7469. Using the mid-IR diagnostic diagrams discussed earlier, we constructed a detailed map identifying the sources of ionization responsible for the observed ratios at each location. However, the question remains of whether the star-forming ring is influenced by the presence of the AGN, and if so, what mechanisms may be responsible.

\begin{figure*}[]
\centering
\includegraphics[width=\textwidth]{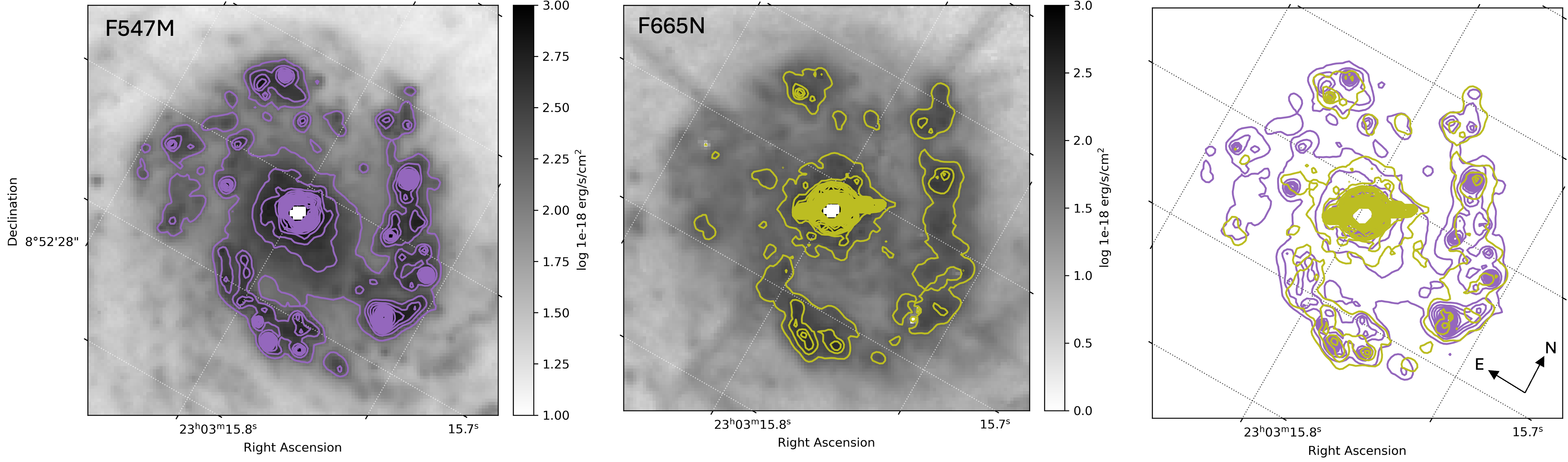}
    \caption{\emph{Hubble Space Telescope} Wide Field Camera 3 images taken with the F547M (continuum) filter and the H$\alpha$ (F665N, continuum subtracted) filter. We over-plot contours on each map to emphasize the morphology of the ring, and plot both sets of contours in the third panel for direct comparison. A readout issue is responsible for the horizontal bar seen at the nucleus of the H$\alpha$ image. The distribution of stars and H$\alpha$ emission-line gas in the eastern part of the ring suggests a disruption of the HII regions by the AGN wind. 
    \label{fig: hubble}}
\end{figure*}

The star formation and AGN models show some overlap in both \sfo based diagnostic diagrams due to intermediate IP lines such as \st and \net being reproducible by either model. However, the overlap occurs only in the most extreme cases of the ionization parameter values for the HII regions' models, and ratios in that overlap region of the diagrams are thus more likely to be indicative of AGN ionizing radiation. The spaxels corresponding to the east wind are consistently overlapping with either or both the AGN and star-forming models, indicating a possible overlap of both ionization sources in that area. However, the question arises whether the wind coming from this AGN would be able to affect the gas at that distance in the first place.

From a purely physical perspective, we can assess whether the AGN at the center of NGC 7469 is capable of ionizing gas at the distance of the ring, potentially impacting the HII regions within it. Using a distance of 542$\pm$98pc for the ring\footnote{Calculated from the F547M image in Figure \ref{fig: hubble}.}, an ionization parameter of 10$^{-3.5}$, and using the Q value calculated earlier, we find that the AGN driving this ULIRG is capable of ionizing gas up to a hydrogen density of approximately 3700$^{+1800}_{-1000}$ cm$^{-3}$. Given that the hydrogen density of a typical HII region is on the order of 10 to a few 100 cm$^{-3}$ \citep{Osterbrock2006}, we would expect the AGN to have a significant impact on the star-forming ring. However, consistent with the findings of \cite{Lai2022}, where they found that the H$_2$/PAH ratios in the ring are within the range of normal photodissociation regions, we detect minimal impact of the AGN on the ring. Indeed, the spaxels associated with the ring consistently fall within the expected values for star-forming regions, except for those oriented towards the east.

The lack of disruption in the star-forming ring, except in one direction, suggests that the ionization cone may only intersect the ring at that specific location. As established, the AGN could disrupt the gas in that region, potentially halting star formation and leading to the noted reduction in \netw emission. Previous studies have found that the \ot distribution around the nucleus is very compact \citep{Fischer2013}, which is consistent with the compactness of lines with similar IP such as \sfo in the MIR (Figure \ref{fig: maps}). Observations of the core of NGC 7469 using the 8.4 GHz Very Large Array (VLA) also revealed symmetrical radio emission around the nucleus and the star-forming ring, with no indication of extended radio jets \citep{Perez-Torres2009}. This suggests that the cone has a narrow opening angle and intersects only a portion of the star-forming ring to the east. The lack of interaction with the ring in the other direction could be due to the other half of the ionization cone being obscured by the disk and the ring, thus affecting mainly the side facing away from us.

NGC 7469 has also been observed by the \emph{Hubble Space Telescope} Wide Field Camera 3 (WFC3) imager using a wide variety of filters, including F665N (ID: 15649, PI: Chandar, exp. time: 2380 s) and F547M (ID: 11661, PI: Bentz, exp. time: 2240 s). These filters allow us to gain insight into the distribution of the stars in the ring and the H$\alpha$ emitting gas respectively. Figure \ref{fig: hubble} shows our reprocessed images using the aforementioned filters with the F665N image having been continuum-subtracted using a scaled version of the F547M image. The continuum subtraction process is as follows: we first used the header information to convert the images to flux units of erg/s/cm$^2$. We next identified off-nuclear regions with strong continuum emission and little evidence of dust or line emission in the F547M continuum band. We then introduced a scaling factor to the continuum band such that the local median values from the continuum filter and the narrow filter are identical in that region, and proceeded with the continuum subtraction. As can be seen, the ring is noticeably clumpy, and the emission is not uniform around the ring. However, the eastern part of the ring is where the greatest difference can be seen between the continuum and the H$\alpha$ distributions, where the H$\alpha$ morphology closely resembles our \netw map (Figure \ref{fig: maps}). In fact, as seen in the third panel of Figure \ref{fig: hubble}, the contours of the two filters match extremely well at all points around the ring, except for the East, where there is a noticeable lack of H$\alpha$ emission. 

At first glance, this discrepancy could be due to three things: extinction, older stars, and disruption of HII regions due to the AGN wind. Extinction, however, cannot be responsible as we are seeing the same structure in the IR, and the H$\alpha$ filter corresponds to a longer wavelength than the continuum, which would extinct the continuum before the H$\alpha$. The second option is that the population in the east part of the ring is simply too old to produce HII regions, leading to the reduction in emission-line gas. However, this is not likely to be the case as it has been shown that the star clusters in the east and the west side of the ring have similar age ranges (10 to 20 Myr, \citealp{Diaz-Santos2007}). Nevertheless, if the east side was on the younger end of this range with the west side on the older end, this could explain some of the discrepancy. The last option would be attributed to the interaction between the AGN wind and the star-formation processes in the east, disrupting the HII regions but not affecting the already-present stellar population. This seems like the most plausible option given the known eastward AGN outflow. 

As demonstrated above, there appears to be some interaction occurring in the direction of the AGN wind. However, it is important to consider whether any large-scale interaction between the AGN and the ring has occurred in the past. For an SMBH of this size (approximately 1.10 x 10$^7$ M$_{\odot}$, \citealp{Lu2021}) at a distance of 542$\pm$98pc, the infall time is 63$^{+18}_{-16}$ Myr. To reiterate, the star-forming ring has been found to consist of both a young stellar population (2 to 5 Myr) and an intermediate-age stellar population with ages approximately 10-30 Myr \citep{Diaz-Santos2007, Bohn2023}, which is consistent with the photoionization model grid used in Section \ref{sec: models}. Since the older stars in the ring appear to be on the same order as the infall time, there may be a connection between the initial burst of star formation and the fueling of the AGN in NGC 7469. However, given the observed lack of large-scale interaction between the AGN and the ring, there is no evidence that the current AGN activity is responsible for initiating the younger burst.

We are thus observing the effect of the AGN outflow on the preexisting stellar population and star formation in the eastern ring. For a wind traveling at 500 km/s, the time required for it\footnote{The value used for the velocity of the wind was derived from the broad wing centroid value of the [Fe \textsc{vii}] 5.4$\mu$m line.} to reach the ring would be approximately 1 Myr. Consequently, there has been sufficient time for the wind to significantly disrupt the gas and thereby disturb the star formation processes in this region. However, to reiterate, this process appears to affect only a small portion of the ring and does not seem to have a large-scale feedback effect \citep{Shimizu2015, Feuillet2024-SFR}, consistent with the findings of \cite{Fischer2017} in Mrk 573, a Seyfert 2 galaxy. To effectively evacuate the bulge, a wider ionization cone would be necessary, which could be generated by a more powerful and more massive AGN. Another option would require the cone to be oriented such that the wind intercepts more of the ISM, coupled with a stronger outflow. A similar analysis of Mrk 477, which hosts a very powerful nuclear starburst \citep{Heckman1997} and has a bolometric luminosity about 10 times that of NGC 7469 \citep{Anna2021}, would provide a valuable comparison for studying feedback mechanisms on sub-kiloparsec scales.

\section{Conclusions}

The close proximity of the star-forming ring to the central AGN in NGC 7469 makes it an ideal target to investigate the starburst-AGN connection in great detail. By using MIRI data to delve into the core of this nearby Luminous Infrared Galaxy, we concluded the following:

\begin{enumerate}
    \item The flux maps of mid-IR emission lines from ions with a range of ionization potentials provide an initial insight into the ionization sources within the core of NGC 7469. These maps distinctly reveal the star-forming ring in both the \netw and \st emission lines, accompanied by a noticeable hole in the flux emission towards the eastern direction, as well as an asymmetry in the fluxes of the higher ionization lines towards the same region. 
    \item Mid-IR diagnostic diagrams such as those using \netnetw plotted against \sfons/\stns, \sfons/\netns, or \nevns/\net are extremely efficient at separating regions of different ionization sources, such as star formation-dominated and AGN-dominated ones down to 100pc scales. These diagrams also effectively separate blue compact dwarf galaxies from other star-forming points, and we find minimal relative contribution from WR stars within the star-forming ring.
    \item We find that NLR Cloudy models using a carefully constructed SED based on X-ray and UV data are in agreement with our AGN-dominated nuclear points. Similarly, the Starburst99 models successfully reproduce the emission line ratios found in the star formation-dominated regions, particularly in the ring. 
    \item The models and comparison to other data suggest that the ionization from the nucleus reaches all the way to the ring, but does not appear to disturb it in any significant way other than in the direction of the east wind. This is possibly due to the orientation of the cone, which only interacts with the ring in the direction of the wind. 
    \item We draw the following conclusions regarding the overall state of the interaction between the AGN and the star-forming ring: (1) The older stellar population may have played a role in the initial fueling of the AGN. (2) The current AGN activity does not appear to be responsible for the formation of the younger stellar population. (3) There is evidence of small-scale negative feedback in the eastern region of the ring.
\end{enumerate}

Although no large-scale feedback has been detected in this case, analysis of the gas dynamics could further support these findings. Future work will focus on the kinematics of this core region to further investigate negative feedback in the eastern part of the ring, as well as determine the x-ray characteristics of the east wind using IR footprint lines \citep{Trindade2022}.


\begin{acknowledgments}
The authors would like to thank the anonymous referee for their valuable comments, which greatly improved this paper. This work is based on observations made with the NASA/ESA/CSA James Webb Space Telescope. The data were obtained from the Mikulski Archive for Space Telescopes at the Space Telescope Science Institute, which is operated by the Association of Universities for Research in Astronomy, Inc., under NASA contract NAS 5-03127 for JWST. These observations are associated with program \#1328. Additionally, this research used observations made with the NASA/ESA Hubble Space Telescope obtained from the Space Telescope Science Institute, which is operated by the Association of Universities for Research in Astronomy, Inc., under NASA contract NAS 5–26555. These observations are associated with programs 11661 and 15649. All of the data presented in this article were obtained from the Mikulski Archive for Space Telescopes (MAST) at the Space Telescope Science Institute. The specific observations analyzed can be accessed via \dataset[doi: 10.17909/djwe-bj23]{https://doi.org/10.17909/djwe-bj23}. We also thank Marina Bianchin for helpful conversations on data calibration.
\end{acknowledgments}

%




\appendix
\section{[O IV] PSF issues} \label{app: PSF}

As mentioned in Section \ref{sec: diags}, \of 25.89$\mu$m is an effective tracer of AGN activity. However, it presents several challenges in the context of the MIRI IFU data. Indeed, \of is very near the long wavelength limit of the MIRI sensitivity of NGC 7469 and the MIRI instrument experiencing increasing fringing issues at longer wavelengths leads to significant fringing around this emission line. \of is also a relatively strong line that is highly concentrated at the nucleus, resulting in point spread function (PSF) issues. Specifically, the \of emission from the inner region spreads into neighboring spaxels, artificially increasing the \of flux away from the nucleus.

\begin{figure*}[h!]
\centering
\includegraphics[width=\textwidth]{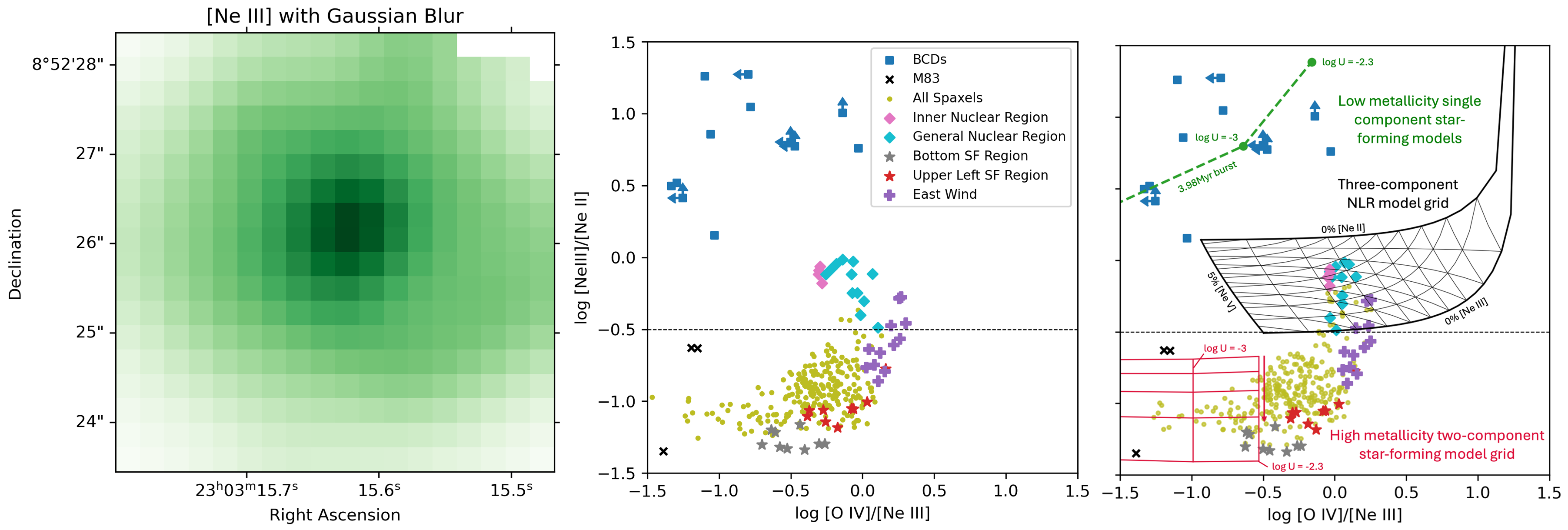}
    \caption{Demonstrating the effect of PSF issues for \of on the resulting diagnostic diagram. Left panel: showing the \net map along with a Gaussian blur applied to simulate the same PSF issue as on the \of map. Center panel: \ofnet diagram without correcting for the PSF issue. The points are defined in the legend, and further details may be found in the caption of Figure \ref{fig: diagnostic diagrams}. Right panel: \ofnet diagram with the PSF issue corrected by applying the Gaussian blur to the \net and \netw maps. The grids correspond to different photoionization models, and more details are included in the caption of Figure \ref{fig: models}. 
    \label{fig: OIV issue}}
\end{figure*}

This phenomenon is evident in the \ofnet distribution in the \cite{Weaver2010} diagram, where one would expect the nucleus to exhibit the highest ratio among the spaxels. However, as shown in Figure \ref{fig: OIV issue}, the shape of the distribution does not align with this expectation, with the wind and spaxels adjacent to the nucleus displaying higher \ofnet ratios than the nucleus itself. This discrepancy arises from the spreading of nuclear \of into adjacent spaxels, a problem not observed with \netns. Given the difficulty of accurately removing the PSF from the \of map, we introduced an artificial PSF to the \net and \netw maps to examine its effect on the diagram \citep{Law2023}. Upon doing so, the distribution of the spaxels aligns more closely with expectations, as now illustrated in the third panel of Figure \ref{fig: OIV issue}.

\bibliography{sample631}{}
\bibliographystyle{aasjournal}



\end{document}